%% file: main.tex
\documentclass[journal]{IEEEtran}

\ifCLASSINFOpdf
\else
   \usepackage[dvips]{graphicx}
\fi
\usepackage{url}

\usepackage{graphicx}
\usepackage{amsmath}
\usepackage{multirow}
\usepackage{booktabs}
\usepackage{amssymb}
\usepackage{subfig}

\begin{document}

\title{UniEnc-CASSNAT: An Encoder-only Non-autoregressive ASR for Speech SSL Models}

\author{Ruchao Fan, \IEEEmembership{Student Member, IEEE}, Natarajan Balaji Shankar, and Abeer Alwan, \IEEEmembership{Fellow, IEEE}
\thanks{This work was supported in part by the NSF.}
\thanks{R. Fan and N. B. Shankar are students with the UCLA ECE Department, California, 90095, USA (e-mail: \{fanruchao,balaji1312\}@g.ucla.edu).}
\thanks{A. Alwan is a Professor with the UCLA ECE Department, Los Angeles, California, 90095, USA (e-mail: alwan@ee.ucla.edu).}}

\markboth{Manuscript submitted to IEEE Signal Processing Letters}
{Shell \MakeLowercase{\textit{et al.}}: Bare Demo of IEEEtran.cls for IEEE Journals}
\maketitle

\begin{abstract}
Non-autoregressive automatic speech recognition (NASR) models have gained attention due to their parallelism and fast inference. The encoder-based NASR, e.g. connectionist temporal classification (CTC), can be initialized from the speech foundation models (SFM) but does not account for any dependencies among intermediate tokens. The encoder-decoder-based NASR, like CTC alignment-based single-step non-autoregressive transformer (CASS-NAT), can mitigate the dependency problem but is not able to efficiently integrate SFM. Inspired by the success of recent work of speech-text joint pre-training with a shared transformer encoder, we propose a new encoder-based NASR, UniEnc-CASSNAT, to combine the advantages of CTC and CASS-NAT. UniEnc-CASSNAT consists of only an encoder as the major module, which can be the SFM. The encoder plays the role of both the CASS-NAT encoder and decoder by two forward passes. The first pass of the encoder accepts the speech signal as input, while the concatenation of the speech signal and the token-level acoustic embedding is used as the input for the second pass. Examined on the Librispeech 100h, MyST, and Aishell1 datasets, the proposed UniEnc-CASSNAT achieves state-of-the-art NASR results and is better or comparable to CASS-NAT with only an encoder and hence, fewer model parameters. Our codes\footnote{\url{https://github.com/Diamondfan/cassnat_asr}} are publicly available.
\end{abstract}


\IEEEpeerreviewmaketitle

\input{Tex/intro}

\input{Tex/method}

\input{Tex/exp_setup}

\input{Tex/results}

\input{Tex/conclusion}


\bibliographystyle{IEEEbib}
\bibliography{refs}

\end{document}

%% file: Tex/intro.tex
\section{Introduction}
\label{sec:intro}

\begin{figure*}[t!]
    \centering
    \subfloat[CASS-NAT]{{\includegraphics[width=0.35\textwidth,height=0.20\textheight]{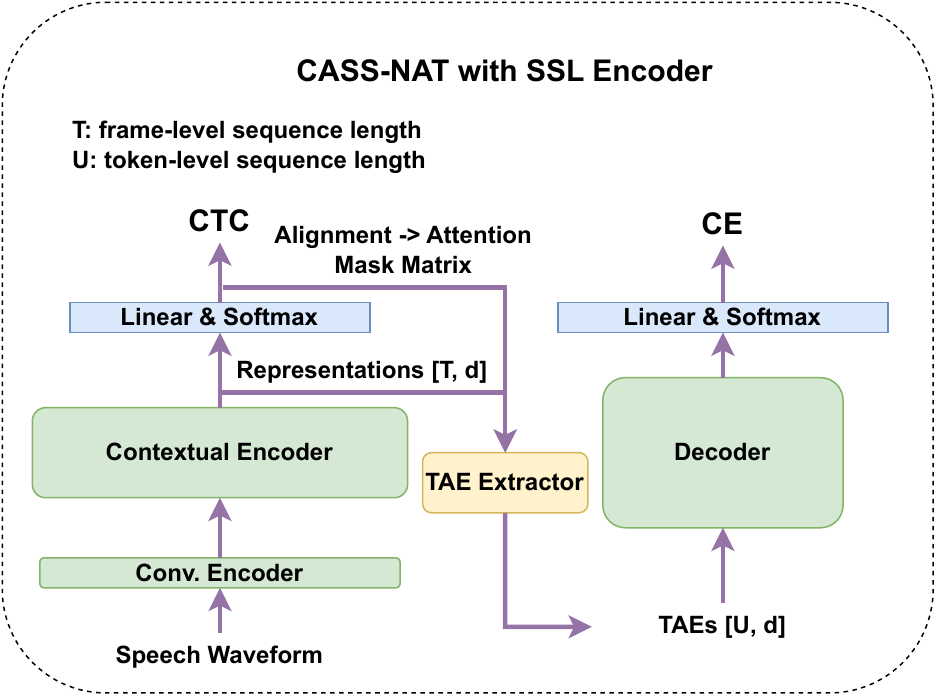} }}%
    \qquad
    \subfloat[UniEnc-CASSNAT]{{\includegraphics[width=0.52\textwidth,height=0.22\textheight]{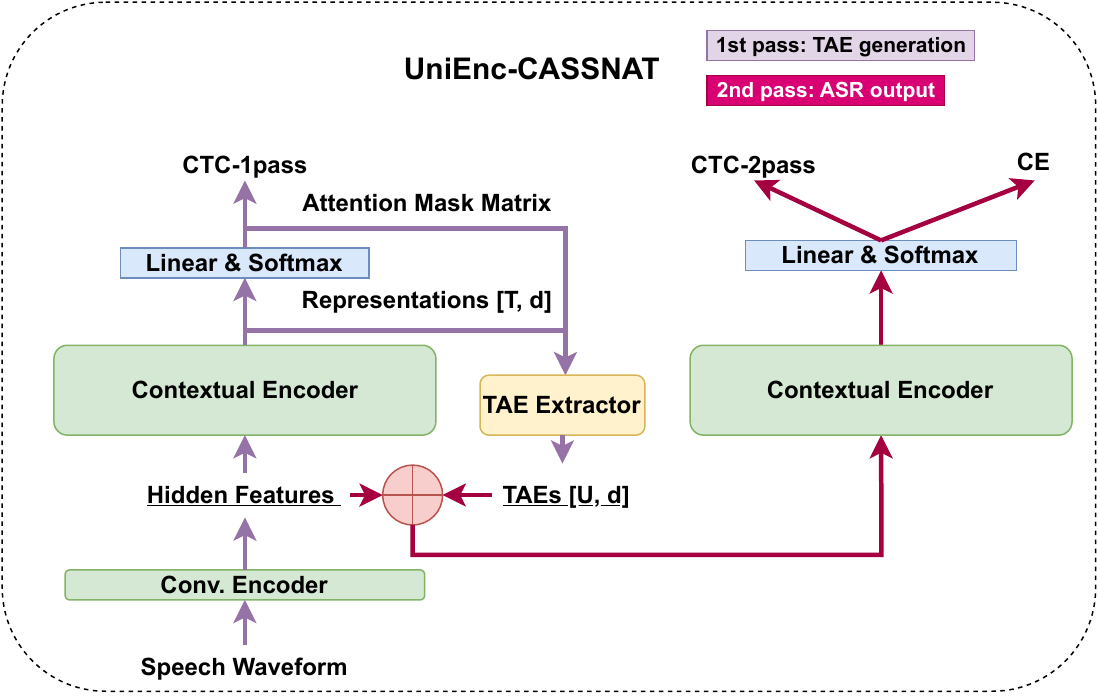}}}%
    \caption{(a): the diagram of CASS-NAT. (b): the proposed UniEnc-CASSNAT. HuBERT conv. and contextual encoders are used. The TAE extractor is a self-attention module that transforms the acoustic representations with length T to TAEs with length U. The generation of TAEs and second pass forward computation are repeated during iterative decoding. }%
    \label{fig:figure}
\end{figure*}

\IEEEPARstart{I}{n} recent years, self-supervised learning (SSL) has become popular in speech \cite{zhang2022bigssl,SSLsurvey2022,chang2021exploration} and natural language \cite{raffel2020exploring,kenton2019bert} processing. The SSL models learn prior knowledge from a large amount of unannotated data and are called pre-trained or foundation models. Widely-used speech foundation models include APC \cite{ChungG20} that predicts future frames from their histories, and Wav2vec2.0 \cite{baevski2020wav2vec}, HuBERT \cite{HsuBTLSM21}, and WavLM \cite{chen2022wavlm} that reconstruct or predict pseudo labels via the masked portions of the speech signal. The speech foundation models are proven effective in improving low-resource tasks by fine-tuning \cite{fan2022towards}.     

Concurrently, non-autoregressive automatic speech recognition (NASR) has attracted considerable interest due to its fast inference \cite{ChenWVZD21,NozakiK21,higuchi2021comparative}. Although it is not naturally designed for streaming ASR, NASR can greatly improve the inference efficiency for offline applications. As the earliest end-to-end ASR framework, connectionist temporal classification (CTC) \cite{graves2006connectionist, ng21b_interspeech} can be regarded as an encoder-based NASR model when using greedy decoding. However, the performance of CTC is always constrained by the output independence assumption. On the other hand, most NASR models are proposed based on the encoder-decoder framework where the decoder can mitigate the output independence problem. For example, the decoder of Mask-CTC \cite{Higuchi0COK20} is a masked language model to correct the low confidence tokens in CTC output. Align-Refine \cite{ChiSK21} uses the decoder to refine the CTC alignment iteratively. LASO \cite{bai2021fast}, CASS-NAT \cite{FanCC021}, and Paraformer \cite{GaoPara21} extract acoustic embedding as the decoder input for token-level contextual representation learning. However, the encoder-decoder framework does not perfectly fit the current foundation models, which are pre-trained with the transformer encoder structure. Although previous work developed pre-trained models \cite{zhang-etal-2022-speechut, ao2022speecht5} for the encoder-decoder framework, it is specifically designed for autoregressive transformers. Additionally, \cite{fan2023ctc} trains the transformer decoder from scratch with the encoder initialized from the speech foundation model. The work in \cite{DengYWHCZ22} and \cite{higuchi-etal-2022-bert} introduce BERT to the NASR model for better output dependency modeling. However, these methods may contain unnecessary model parameters.

In this work, based on previous method (CASS-NAT) \cite{FanC00A21}, we present a new encoder-only NASR (UniEnc-CASSNAT) that can function in a way that is similar to CASS-NAT encoder and decoder. Like CTC, UniEnc-CASSNAT can be initialized from speech foundation models (HuBERT base model \cite{HsuBTLSM21} is used). To behave as both the CASS-NAT encoder and decoder, UniEnc-CASSNAT has two forward passes and accepts two types of input for each. In the first pass, speech features (output of HuBERT Conv. encoder) are fed into the contextual encoder to generate token-level acoustic embeddings (TAEs). In the second pass, the concatenation of speech features and the TAEs (along the time dimension) are used as the contextual encoder inputs. The TAE corresponding outputs are selected for ASR loss computation. The outputs in the second pass can generate better quality TAEs than those in the first pass and hence lead to better ASR performance. We, therefore, further propose a multi-pass CTC (MP-CTC) training and iterative decoding method to improve the WER performance. Experiments on Librispeech 100-hour \cite{panayotov2015librispeech}, MyST \cite{ward2011my}, and Aishell1 \cite{bu2017aishell} datasets show that the proposed methods can achieve better or comparable WERs to CASS-NAT, and contain fewer parameters. The framework can be applied to other encoder-decoder-based NASR.

The remainder of this paper is organized as follows. Section \ref{sec:method} introduces the framework of UniEnc-CASSNAT and iterative decoding process. Experimental setups are described in Section \ref{sec:exp_setup}. Results are shown and discussed in Section \ref{sec:results}. We conclude the paper in Section \ref{sec:conclusion}.


%% file: Tex/method.tex
\section{Proposed Framework: UniEnc-CASSNAT}
\label{sec:method}


\subsection{Encoder-Decoder CASS-NAT}
\label{ssec:cassnat}

CASS-NAT \cite{FanCC021} consists of an encoder, a token-level embedding extractor (TAEE), and a decoder as plotted in Figure \ref{fig:figure}(a). The connectionist temporal classification (CTC) \cite{graves2006connectionist} loss is added to learn the alignment between the acoustic and token sequences. The alignment can provide segmentation information for each token. TAEE extracts an embedding for each token from encoder outputs (with a shape of $[T,d]$, where $T$ is the frame length and $d$ is the hidden dimension) using the segmentation information. The extracted token-level acoustic embeddings (TAEs) (with a shape of $[U,d]$, where $U$ is the token sequence length) are fed into the decoder, which models the relationship between tokens. Suppose the input sequence is $X=\{x_1,...,x_t,...,x_T\}$, the ground truth is $Y=\{y_1,...y_u,...,y_U\}$ and the CTC alignment is $Z$, then the objective function on the decoder side can be written as:
\begin{equation}
\label{eq:maxapprox}
\begin{aligned}
   L_{\text{dec}} &= \log P(Y|X) \\
   &\geq \mathbb{E}_{Z|X}[\log P(Y|Z,X)] \\
   &\approx \max_Z{\log\prod_{u=1}^U{P(y_u|z_{t_{u-1}+1:t_{u}}, x_{1:T})}}
\end{aligned}
\end{equation}

where $t_{u-1}+1:t_{u}$ represents the acoustical boundary for token $u$ provided by the alignment $Z$. We use a maximum approximation for the expectation (Viterbi-alignment during training). CASS-NAT is then trained by jointly maximizing the decoder loss in Eq. \ref{eq:maxapprox} and the CTC loss on the encoder side with a task ratio $\lambda$.
\begin{equation}
\label{eq:joint}
    L_{\text{joint}} = L_{\text{dec}} + \lambda \cdot \log \sum_{Z\in q}{\prod_{i=1}^T{P(z_i|X)}}
\end{equation}
where $q$ is all the alignments that can be mapped to the label $Y$ by removing blank tokens in CTC and repetitions. 

During decoding, Viterbi-alignment is not available. We therefore use error-based sampled alignments (ESA) (see details in~\cite{FanCC021}), where the multiple alignments $Z$ are sampled based on the CTC greedy search output with low confidence scores. The TAEs computed from the sampled alignments are fed into the decoder to obtain multiple ASR outputs. The autoregressive transformer provides a ranking score for each ASR output (one ASR output corresponds to one alignment).

\subsection{Encoder-only CASS-NAT: UniEnc-CASSNAT}
\label{ssec:unienc-cassnat}
Speech foundation models are proven to be useful in downstream ASR tasks. The encoder of CASS-NAT can be inherited from a speech foundation model and extracts better acoustic representations~\cite{fan2023ctc}. However, the CASS-NAT decoder has to be trained from scratch. Inspired by the success of recent work of speech-text joint pre-training~\cite{tang-etal-2022-unified,wang2023understanding} with a shared encoder, we rethought the necessity of CASS-NAT decoder and propose an encoder-only CASS-NAT, denoted as UniEnc-CASSNAT to fit the size of the speech foundation models.

The UniEnc-CASSNAT is shown in Figure \ref{fig:figure}(b) with two forward passes. In the first pass, the hidden features extracted from the conv. encoder are fed into the contextual encoder for CTC modeling and the token-level acoustic embeddings (TAEs) are extracted using the alignment information from CTC outputs. In the second pass, the extracted TAEs ($[U,d]$) are concatenated with the hidden features ($[T,d]$) (along the time dimension) to be the input to the contextual encoder. The self-attention layer in the contextual encoder enables frame-frame, frame-token, and token-token interactions between hidden features and TAEs. Note that the goal of the first pass is to obtain TAEs, whose quality is highly related to the ASR performance. The better the speech foundation model, the better the quality of the TAEs extracted by UniEnc-CASSNAT. The second pass is similar to the role of the CASS-NAT decoder for modeling the relationships between TAEs and frame-level hidden features. We investigate whether the encoder is capable of both frame-level acoustic representation learning and contextual modeling between tokens.

\subsection{MP-CTC Training and Iterative Decoding}
\label{ssec:iterative_decode}
The output of the second pass is a sequence of $T+U$ vectors, where the first $T$ vectors correspond to hidden features, and the $U$ vectors correspond to TAEs. Since the quality of TAEs is essential to the performance of the CASS-NAT decoder, we propose to add another CTC loss to the first $T$ outputs of the second pass and formulate a multi-pass CTC (MP-CTC) training. With the CE loss used on the $U$ outputs, the final objective function of UniEnc-CASSNAT can be written as:
\begin{equation}
\label{eq:multictc}
    L_{\text{unienc-cassnat}} = L_{\text{dec}} + \lambda_1 \cdot L_{CTC-1pass} + \lambda_2 \cdot L_{CTC-2pass}
\end{equation}

We share the final feed-forward layer for the two CTC losses. Theoretically, the second-pass CTC loss would have better performance than the first pass because it accepts additional input information (TAEs). An intuitive idea is to iteratively improve the quality of TAEs by repeating the second pass with newly extracted TAEs. Hence, we propose an iterative decoding method for UniEnc-CASSNAT. Specifically, we define the hidden features as $H$, and the first pass of UniEnc-CASSNAT encoder as $\text{Iter}_{0}$. $\text{Iter}_{0}$ would generate $\text{TAE}_0$. The second pass uses $H + \text{TAE}_0$ as input and generates $\text{TAE}_1$, which we define as $\text{Iter}_{1}$. Generally, for iteration $n$, the contextual encoder accepts $H$ and $\text{TAE}_{n-1}$ as input and generates $\text{TAE}_n$ for the iteration $n+1$. In each iteration, ESA generates multiple TAEs for the next iteration. We define the number of sampled alignments in each iteration as $S_n$. The total number of the sampled alignments would be $\prod_{n=0}^{N-1} S_n$, where $N$ is the number of iterations used in the decoding. We empirically found that two iterations are sufficient for a desirable word error rate (WER).

%% file: Tex/exp_setup.tex
\section{Experimental Setup}
\label{sec:exp_setup}

\subsection{Data Settings}
\label{ssec:dataset}
The experiments were conducted on three datasets: the 100-hour subset of LibriSpeech English corpus \cite{panayotov2015librispeech}, the 240-hour (annotated section) My Science Tutor (MyST) children's speech corpus \cite{ward2011my}, and 170-hour Aishell1 Mandarian corpus \cite{bu2017aishell}. We chose the 100-hour subset of Librispeech to enable comparisons with previous work on NASR. We conducted pre-processing on MyST dataset to get a better baseline compared to \cite{fan2023ctc}. For example, we mapped filling pauses, non-speech events, and truncated words to $\langle\text{UNK}\rangle$. The $\langle \text{UNK}\rangle$ is not considered when computing WER.

The sets of output labels consist of 1024 word-pieces for Librispeech 100h and 500 word-pieces for the MyST, obtained by the SentencePiece method \cite{kudo-richardson-2018-sentencepiece}. For Aishell, 4230 characters are used as the vocabulary.

\subsection{Model Settings}
\label{ssec:model_setup}
A CTC/Attention autoregressive transformer (AT) baseline was first trained with an architecture of a 12-block encoder and a 6-block decoder. Suppose the tuple of a transformer setting is represented by (model dimension, feed-forward layer dimension, number of heads in self-attention), we define three settings: $d_{512}$ for (512, 2048, 8), $d_{768}$ for (768, 3072, 12), and $d_{256}$ for (256, 2048, 4). $d_{512}$ is used for the two English datasets, and $d_{256}$ is used for the Aishell1 dataset. Later on, we follow the same setting as in \cite{fan2023ctc} for CASS-NAT training. For a fair comparison, we also include a CTC baseline as an encoder-only NASR architecture. When training with the speech foundation models, the 12-block encoder was replaced with a HuBERT-base model, either the English\footnote{\url{https://dl.fbaipublicfiles.com/hubert/hubert_base_ls960.pt}} version for Librispeech and MyST, or the Chinese version\footnote{\url{https://huggingface.co/TencentGameMate/chinese-hubert-base}} for Aishell1. We also conducted experiments on the TAE extractor in UniEnc-CASSNAT to examine the trade-off between performance and model size.

All models are optimized using a noam scheduler \cite{vaswani2017attention} with warmup steps of 15k (10k for Librispeech 100h), a peak learning rate of 5e-5 for the encoder, and 1e-3 (5e-4 for MyST) for uninitialized modules. The models were trained using a batch size of 80s audio samples (40s for MyST because it contains longer utterances). The training either stops when the WER of the valid set doesn't improve for 10 epochs or is terminated at 30 epochs. For MP-CTC training, the task ratio of CTC loss in the second pass is set to one.

All results are decoded without the usage of the external language model. For the AT baseline, the beam search decoding is applied with a beam size of 20 for Librispeech and MyST, and 10 for Aishell1. For CASS-NAT, the number of sampled alignments is 50 and the threshold is 0.9. We explore the effects of the number of sampled alignments in two iterations, and the threshold for each iteration is set to 0.9 as well.

%% file: Tex/results.tex
\section{Results and Discussion}
\label{sec:results}

\subsection{Main Results}
\label{ssec:main_results}

\input{Tables/main_results}
The main WER results of UniEnc-CASSNAT on the Librispeech 100h, MyST, and Aishell1 datasets are shown in Table \ref{tab:main_results}. We first train two autoregressive transformer baselines with or without the usage of self-supervised learning. The results in the table again show the effectiveness of the speech foundation models. CASS-NAT achieves close performance to their AT counterpart, which is consistent with previous work. We also present the results of CTC on the three datasets. Due to the output-independent assumption, CTC is worse than the AT baseline and CASS-NAT although it requires fewer parameters. Note that the motivation of UniEnc-CASSNAT is to investigate whether the encoder can jointly model the frame-level and token-level acoustic embedding without the use of the decoder and thus has fewer model parameters. We expect to obtain a model with similar model parameters compared to CTC but close performance to the CASS-NAT. As shown in Table~\ref{tab:main_results}, the proposed UniEnc-CASSNAT achieves comparable or better results than CASS-NAT, for example, a WER of 11.0\% for UniEnc-CASSNAT vs. 11.2\% for CASS-NAT on the Librispeech test-other set, but is superior to CASS-NAT in terms of model size (99.3M vs. 130.5M). A smaller model size can be helpful for on-device deployment. Compared to CTC, the UniEnc-CASSNAT achieves much better performance than CTC with a similar model size. The additional 3M parameters compared to CTC (95.7M) are from the TAE extractor. The limitation of UniEnc-CASSNAT could be its slower inference than CTC and CASSNAT because of the multiple forward computations of the encoder with a longer input sequence (concatenation of frames and tokens). The RTF values in Table~\ref{tab:main_results} show that the UniEnc-CASSNAT is still 3-5x faster than the AT models although it is 6x slower than CASS-NAT.

The proposed UniEnc-CASSNAT achieves the best-performing NASR results so far in the literature \cite{fan2023ctcbert,fan2023ctc} on Librispeech 100h and MyST. One can find better WER performance on the Librispeech 100h data, for example, in \cite{zhang-etal-2022-speechut,wu2023wav2seq}. However, in that work, the authors either use a larger model trained with Libri60k hours of data or extra text data. We compare the UniEnc-CASSNAT results to a similar work BERT-CTC~\cite{higuchi-etal-2022-bert}, which also uses an encoder-only structure. Differently, UniEnc-CASSNAT generates ASR outputs with CE loss instead of CTC loss in BERT-CTC and does not require a pre-trained BERT module (smaller in size than BERT-CTC). In addition, UniEnc-CASSNAT achieves the best performance with two iterations only instead of more than 10 iterations in BERT-CTC (faster inference). Based on the results in Table~\ref{tab:main_results}, UniEnc-CASSNAT is better on Librispeech-100h but worse on Aishell1 than BERT-CTC. The reason could be that Aishell1 contains simple sentences where a pre-trained BERT model is more beneficial ~\cite{higuchi-etal-2022-bert} and the BERT-CTC has 143M parameters versus 102M in UniEnc-CASSNAT.

\subsection{Ablation Study of UniEnc-CASSNAT}
\label{ssec:unienc-cassnat-study}
We present more results on the Librispeech 100h data to show the importance of the proposed MP-CTC training and iterative decoding. First, we set $\lambda_2$ in Eq. \ref{eq:multictc} to zero and train a UniEnc-CASSNAT with only first-pass CTC. The results in Table \ref{tab:study} show that the single-pass CTC (SP-CTC) training has a performance gap compared to the CASS-NAT. Additionally, SP-CTC training is not able to perform iterative decoding because $\text{TAE}_{n>1}$ is not constrained by CTC outputs. MP-CTC training is also worse than the CASS-NAT without iterative decoding (e.g. (50, NA)). When applying iterative decoding, we explore different combinations of the number of sampled alignments $S_n$ in each iteration. The total number of sampled alignments is set to the same as that used in CASS-NAT for a fair comparison. As shown in Table \ref{tab:study}, iterative decoding with a setting of $(25, 2)$ achieves the best WER performance and is better than the WER of CASS-NAT. Most of the combinations of $S_n$ achieve comparable WERs to CASS-NAT. It is also noted that the diversity of sampled alignments in the first iteration is more important than that in the second iteration.

Finally, since the TAE extractor introduces extra model parameters besides the foundation model, we conduct experiments of UniEnc-CASSNAT with different transformer settings ($d_{256}$, $d_{512}$, $d_{768}$) described in Section \ref{ssec:model_setup}). The results are also shown in Table \ref{tab:study}. We can see from the table that with a bigger TAEE module, the performance tends to be better. However, we select MP-CTC-$d_{512}$ as the final results to show in Table \ref{tab:main_results} because MP-CTC-$d_{768}$ did not achieve significant improvements with additional 5M parameters.

%% file: Tables/main_results.tex
\begin{table*}[t]
\centering
\caption{WER performance of UniEnc-CASSNAT and comparisons to previous methods on Librispeech-100h, MyST, and Aishell1 datasets. State-of-the-art (SOTA) results with the usage of speech foundation models that are pre-trained with the same amount of unannotated data to ours are reported. The real-time factor (RTF) of each method on Librispeech test-other data is presented for speed comparison. All bold-faced improvements are statistically significant. }
\begin{tabular}{l c c c c c c c c c c c}
\hline
\multirow{2}{*}{Model Type} & \multirow{2}{*}{Model Size}&\multicolumn{5}{c}{Librispeech-100h} & \multicolumn{2}{c}{MyST} & \multirow{2}{*}{Model Size} & \multicolumn{2}{c}{Aishell1} \\
\cmidrule(r){3-7} \cmidrule(r){8-9} \cmidrule(r){11-12}
 & & dev-clean & dev-other & test-clean & test-other & RTF & dev & test & & dev &test \\ 
\hline\hline
\rule{0pt}{2ex}
AT-w/o SSL & 85.1M & 6.6 &18.2 &6.9 &18.2 & 0.325 & 13.5 &14.9 & 33.6M & 4.6 & 5.0 \\
\cline{1-12}
\rule{0pt}{2ex}
AT-w/ SSL &121.6M &4.8 &11.0 &4.8 &10.8 & 0.486 & 11.4 &13.1 &107.3M &4.0 &4.3 \\
\hline
\multicolumn{12}{c}{Non-autoregressive ASR} \\
\hline\hline
\rule{0pt}{2ex}
Previous SOTA & w/ SSL & 4.6 & 11.3 & 4.8 & 11.3\textsuperscript{\cite{fan2023ctcbert}} & - & 16.0 & 15.6\textsuperscript{\cite{fan2023ctc}} & w/ SSL & 3.6 & 3.8\textsuperscript{\cite{zheng2021wav}} \\
\hline
\rule{0pt}{2ex}
BERT-CTC~\cite{higuchi-etal-2022-bert} &- & 7.0& 16.3 & 7.2 & 16.6 & - & - & - & 143M & 3.9 & 3.9\\
\hline
\rule{0pt}{2ex}
 CTC &95.7M & 6.1& 13.8 & 6.2 & 13.8 & 0.005 & 12.9& 14.5 & 95.7M & 4.5& 4.9\\\hline
\rule{0pt}{2ex}
CASS-NAT &130.5M &\textbf{4.7} &11.4 &4.9 &11.2 & 0.014 & 11.9 &\textbf{13.5} &109.7M &\textbf{4.0} &\textbf{4.3} \\\hline
\rule{0pt}{2ex}
UniEnc-CASSNAT &99.3M &4.9 &\textbf{11.0} &\textbf{4.8} &\textbf{11.0} & 0.093 &\textbf{11.8} &\textbf{13.5} & 102.7M &4.2 & 4.5\\ 
\hline
\end{tabular}
\label{tab:main_results}
\end{table*}

\begin{table}[!htp]\centering
\caption{Ablation study of MP-CTC training, the size of the TAE module, and the iterative decoding. $d_{256}$, $d_{512}$, and ${d_{768}}$ are defined in Section \ref{ssec:model_setup} and their model sizes (including encoder) are 96.1M, 99.3M, 104.2M, respectively. $S_n$ is the number of sampled alignments in the iteration $n$.}\label{tab:study}
\scriptsize
\begin{tabular}{|l|c|c|c|c|c|}
\hline
\rule{0pt}{2ex}
Model Type & $(S_1,S_2)$ &dev-clean &dev-other &test-clean &test-other \\ \hline
CASS-NAT &(50, NA) &4.7 &11.4 &4.9 &11.2 \\ \hline
&\multicolumn{4}{c}{UniEnc-CASSNAT}& \\ \hline
\rule{0pt}{2ex}
SP-CTC &(50, NA) &4.9 &11.9 &5.0 &11.6 \\ \hline
\multirow{6}{*}{MP-CTC-$d_{512}$} &(50, NA) &5.0 &11.7 &5.2 &11.8 \\ \cline{2-6}
&(50, 1) &5.0 &11.1 &4.9 &11.1 \\ \cline{2-6}
&(25, 2) &\textbf{4.9} &\textbf{11.0} &\textbf{4.8} &\textbf{11.0} \\ \cline{2-6}
&(10, 5) &4.9 &11.1 &4.9 &11.1 \\ \cline{2-6}
&(5, 10) &4.9 &11.1 &4.9 &11.2 \\ \cline{2-6}
&(2, 25) &5.0 &11.4 &5.1 &11.4 \\ \cline{2-6}
&(1, 50) &5.2 &11.5 &5.3 &11.6 \\ \hline
MP-CTC-$d_{256}$ &(25, 2) &4.9 &11.4 &4.9 &11.2 \\ \hline
MP-CTC-$d_{768}$ &(25, 2) &4.7 &11.2 &4.8 &11.0 \\ \hline
\end{tabular}
\end{table}

%% file: Tex/conclusion.tex
\section{Conclusion}
\label{sec:conclusion}

We present a novel encoder-only non-autoregressive ASR (NASR) model, UniEnc-CASSNAT, which integrates the advantage of CTC and CASS-NAT. The encoder of UniEnc-CASSNAT acts as both the encoder and decoder in CASS-NAT to reduce the model parameters and can be well initialized from the speech foundation models. Furthermore, MP-CTC training and iterative decoding are proposed for UniEnc-CASSNAT to further improve the performance to be better or comparable to CASS-NAT. We examined the effectiveness of the proposed methods on the Librispeech 100h, MyST, and Aishell1 datasets. To the best of our knowledge, we have achieved the best-performing WER results for NASR on the first two datasets with the same settings as those in the literature. Future work includes model compression and distillation to further reduce the parameters.  